\documentclass[twocolumn,showpacs,preprintnumbers,amssymb]{revtex4}

\usepackage{graphicx}
\usepackage{epsfig}
\usepackage{dcolumn}
\usepackage{bm}

\def\gsim{\lower0.5ex\hbox{$\:\buildrel >\over\sim\:$}}
\def\lsim{\lower0.5ex\hbox{$\:\buildrel <\over\sim\:$}}

\def \rp{R_P}
\def \lp{{L\!\!\!/}}

\begin{document}

\title{Effective R-parity violation from supersymmetry breaking}

\author{Shaouly Bar-Shalom}
\email{shaouly@physics.technion.ac.il}
\author{Sourov Roy}
\email{roy@physics.technion.ac.il}
\affiliation{Physics Department, Technion-Israel Institute of Technology, 
Haifa 32000, Israel}
\date{\today}

\begin{abstract}
We present a scenario in which Yukawa-like R-parity violating (RPV) 
couplings are naturally suppressed.
In our model, RPV is assumed to originate from the SUSY breaking 
mechanism 
and then transmitted into the SUSY Lagrangian 
only through soft SUSY breaking operators in the scalar potential. 
The RPV Yukawa-like 
operators of the superpotential, conventionally parametrized by
the couplings $\lambda,~\lambda^\prime$ and 
$\lambda^{\prime \prime}$, 
are then generated through loops containing the SUSY scalars, 
the gauginos and the soft RPV interactions and are, therefore, 
manifest as effective operators 
with a typical strength of ${\cal O}(10^{-3})$.
\end{abstract}


\maketitle

One of the unresolved puzzles associated with supersymmetric (SUSY) theories 
is whether or not R-parity ($\rp$) is conserved in the SUSY Lagrangian. 
On the one hand, a general SUSY theory allows the existence of 
RPV
trilinear operators in the superpotential 
whose strength is naturally of ${\cal O}(1)$. On the other hand, 
experimental 
searches for RPV interactions yield null results, which 
seem to indicate that such RPV couplings, if exist, are much 
smaller than ${\cal O}(1)$ as long as the typical 
SUSY mass scale is not much larger than the electroweak scale.    
The question then arises: why are the RPV Yukawa-like interactions 
so suppressed wherever they can emerge? 

In this paper we propose a way for generating 
RPV Yukawa-like effective operators, with a structure similar 
to the ones which may appear in the superpotential 
(i.e., the usual $\lambda,~\lambda^\prime$ and $\lambda^{\prime \prime}$ type 
RPV operators) \cite{RPVreview}, however, with a typical 
strength $<<{\cal O}(1)$. We assume that the superpotential conserves $\rp$
and that RPV ``leaks'' into the SUSY Lagrangian through the mechanism 
that breaks supesymmetry. In particular, in this scenario, 
the only place where RPV interactions show up  
is in the scalar potential - in the form of three-scalar ``soft'' operators.  
Being ``soft'', such scalar RPV operators cannot renormalize the Yukawa-like 
RPV operators  and, so, {\it the latter remain zero at all scales}. 
Such a non-vanishing low-energy RPV soft operator will, however, 
generate an effective RPV Yukawa-like interaction through loops 
involving the soft interactions (for a related work see \cite{burzumaty}). 
In general, one expects that these effective
RPV interactions will show additional patterns (compared to the 
``regular'' $\lambda,~\lambda^\prime$ and $\lambda^{\prime \prime}$ ones) due
to their explicit dependence on the particles that run in the loops. 

To illustrate how the new effective RPV couplings are generated,
assume that $\rp$ is violated in the Lagrangian 
{\it only} through the pure leptonic (lepton number violating) 
soft SUSY breaking operators:

\begin{eqnarray}
\Delta V_{\rp,\lp}^{soft} = \epsilon_{ab} \frac{1}{2}
a_{ijk} {\tilde L}^a_i {\tilde L}^b_j {\tilde E}_k^c + {\rm h.c.}~,
\label{slp}
\end{eqnarray}  

\noindent where   
${\tilde L}({\tilde E}^c)$ are the scalar components 
of the leptonic SU(2)
doublet(charged singlet) supermultiplets ${\hat L}({\hat E}^c)$, 
respectively, 
$\tilde L = (\tilde\nu_L, \tilde e_L)$ and 
${\tilde E}^c= \tilde e_R$. Also, 
$a_{ijk}=-a_{jik}$, 
due to the SU(2) indices $a,~b$.

This obviously means that in our framework the superpotential conserves 
$\rp$, i.e., $\lambda$, $\lambda^\prime$ and $\lambda^{\prime \prime}$ as 
well as the bilinear RPV terms ($\mu_i \hat L_i \hat H_u$) are absent. For 
example, $\lambda_{ijk} \to 0$ in 

\begin{eqnarray} 
\Delta W_{\rp,\lp} = \epsilon_{ab} \frac{1}{2}
\lambda_{ijk} {\hat L}^a_i {\hat L}^b_j {\hat E}_k^c  + {\rm h.c.}~.
\label{lp} 
\end{eqnarray} 

\noindent In order to realize this scenario, let us suppose that SUSY 
breaking occurs spontaneously in a hidden sector at the intermediate scale 
$\Lambda \sim 10^{10}-10^{11}$ GeV, described by the $\rp$-conserving 
Fayet-O'Raifearartaigh superpotential:

\begin{eqnarray}
W=m_{12}\hat\Phi_1\hat\Phi_2+g\hat\Phi_3\left( \hat\Phi_2^2-M^2 \right) 
\label{FOM}~,
\end{eqnarray}

\noindent where $m_{12} \sim M \sim \Lambda$ and under $\rp$ 
the chiral superfields in (\ref{FOM}) transform as: 
$\hat\Phi_1,\hat\Phi_2,\hat\Phi_3 \to -\hat\Phi_1,-\hat\Phi_2,\hat\Phi_3$. 
SUSY breaking can then be triggered by 
the vacuum expectation values (VEV's) of the 
auxiliary F-term ($F_{\Phi_i}$) of $\hat\Phi_1,\hat\Phi_2,\hat\Phi_3$.   
We choose the minimum of the potential in this model to lie at 
$A_{\Phi_1}=A_{\Phi_3}=0$, $A_{\Phi_2}=M \sqrt{1-m_{12}^2/(2g^2M^2)}$, which 
satisfies $F_{\Phi_1}=m_{12}A_{\Phi_2}$, $F_{\Phi_2}=0$ and 
$F_{\Phi_3}=g(A_{\Phi_2}^2-M^2)$. 

Supergravity mediation of SUSY breaking can then be parametrized 
by the following $\rp$-conserving superpotential:

\begin{eqnarray}
{\frac 1 {M_{Pl}}} \int d^2 \theta \left[ {\hat \Phi_1}{\hat L}{\hat L}{\hat
E^c} + {\hat \Phi_2}{\hat L}{\hat L}{\hat E^c} \right] 
+ h.c. \label{sup}~,
\end{eqnarray} 

\noindent which will spontaneously break $\rp$ and induce 
the soft operator in (\ref{slp}) with 
$a \sim {\frac {F_{\Phi_1}} {M_{Pl}}} \sim 
{\frac {\Lambda^2} {M_{Pl}}}  \sim m_W$, for 
$m_{12} \sim M \sim \Lambda$. The superpotential in (\ref{sup})
will also generate the operators 
$\propto \lambda$ in (\ref{lp}) with an extremely suppressed 
coupling: $\lambda \propto \frac{A_{\Phi_2}}{M_{Pl}} \sim 
{\frac {\Lambda} {M_{Pl}}} \sim 10^{-9}-10^{-8}$ at the high scale, 
essentially causing the soft operator in (\ref{slp}) to be 
the only source for RPV in this model.
Alternative mechanisms for the 
generation of RPV were suggested in \cite{related}.

In general, other operators which couple the hidden sector 
superfields $\hat\Phi_1,\hat\Phi_2,\hat\Phi_3$ 
to the observable sector can be constructed, which will generate 
an RPV $\mu$-like term as well as a soft bilinear term 
$B_i \tilde L_i H_u$ (for a somewhat similar example see e.g., \cite{liu}). 
One can always rotate away the $\mu_i$
term at the high scale by a field redifinition. In the absence of the
Yukawa-like trilinear terms, $\mu_i$ will remain zero at all scales
at the one loop order (see \cite{carlos}). The $B_i$'s can also be rotated 
away at the high scale, however, even if they are zero at the high scale 
they will be radiatively (one-loop) generated by the non-zero soft trilinear 
$a$ term in (\ref{slp}) and evolved down to the electroweak (EW) scale 
through the RGE. Since, for $\lambda \to 0$ and $\mu_i \to 0$ the RGE for 
$B_i$ takes a relatively simple form \cite{carlos}, $B_i(M_Z)$ can be easily 
estimated. In the leading log we get:

\begin{equation}
B_i(M_Z) \sim -\frac{1}{16\pi^2} h_\tau(M_Z) \mu(M_Z) a_{i33}(M_Z) 
\ln \left(\frac{M_{Pl}^2}{M_Z^2}\right) \label{bi}~,
\end{equation}

\noindent where $h_\tau$ is the $\tau$ Yukawa coupling and $\mu$
is the usual $\mu$ term ($\mu \hat H_u \hat H_d$).

%
%

\begin{figure}
\includegraphics{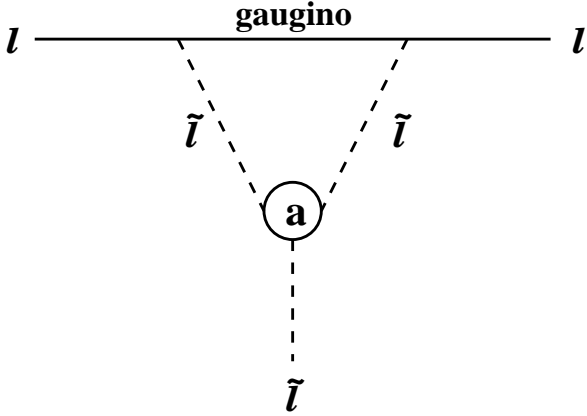}
\caption{\label{fig1} A typical one-loop diagram that generates 
an effective $\lambda$-like operator similar to (\ref{lp}). 
$\tilde\ell$ is either a charged slepton 
or a sneutrino and $\ell$ is a lepton or a neutrino. $a$ is
the trilinear soft breaking coupling defined in (\ref{slp}).}
\end{figure}

Let us now define the effective RPV terms which are generated 
at one-loop through the soft operator in (\ref{slp}) 
via diagrams 
of the type shown in Fig.~\ref{fig1} as follows:

\begin{widetext}
\begin{eqnarray}
{\cal L}_{\rp}^{eff} \equiv 
\frac{1}{2} \left( \frac{a_{ijk}}{16 \pi^2} \right) \left[ 
\tilde\nu_{Li} {\bar e}_k \left( A_{L,ijk} L + A_{R,ijk} R \right) e_j 
+  B_{L,ijk} \tilde\nu_{Li} {\bar\nu}_k L \nu_j 
+C_{L,ijk} {\tilde e}_{Lj} {\bar e}_k L \nu_i
+ D_{L,ijk} {\tilde e}_{Rk} {\bar\nu}_i^c  L  e_j
- (i \to j) \right] +{\rm h.c.},
\label{effrpv}
\end{eqnarray}
\end{widetext}

\noindent where $i,j,k$ are generation indices and 
$L(R) \equiv (1-(+)\gamma_5)/2$. 
Note that the customary RPV operators in 
(\ref{lp}) are obtained by setting: 
$A_{R,ijk}=B_{L,ijk}=0$,
$A_{L,ijk}=C_{L,ijk}=D_{L,ijk}=1~{\rm GeV}^{-1}$ and 
$a_{ijk}=16 \pi^2 \lambda_{ijk}$ GeV.

The particles running in the loop diagrams for each of the form factors 
in (\ref{effrpv}) are: 
(I) ${\tilde e}_L {\tilde e}_R \chi^0_n$ for the 
$\tilde\nu_{Li} {\bar e}_k e_j$ term, where $\chi_n^0$ are the 
four neutralinos ($n=1-4$), (II) ${\tilde e}_L {\tilde e}_R \chi_m$ for the 
$\tilde\nu_{Li} {\bar\nu}_k \nu_j$ term, where $\chi_m$ are the 
two charginos ($m=1,2$), (III) ${\tilde e}_R {\tilde\nu}_L \chi^0_n$ for the 
${\tilde e}_{Lj} {\bar e}_k \nu_i$ term and (IV) both 
${\tilde e}_L {\tilde\nu}_L \chi^0_n$ and 
${\tilde e}_L {\tilde\nu}_L \chi^m$ for the 
${\tilde e}_{Rk} {\bar\nu}_i^c e_j$ term.

Calculating these loop diagrams we find that 
$B_{L,ijk} \propto m_\nu$, i.e., 
no ${\tilde\nu} \bar\nu \nu$ term in the limit of zero neutrino 
masses. Also, $A_{R,ijk} \ll A_{L,ijk}$ 
since $A_{R,ijk}$ is proportional to the 
leptonic Yukawa couplings. Thus, 
neglecting terms which are proportional to the external lepton masses 
and to the leptonic Yukawa couplings we find 

\begin{widetext}
\begin{eqnarray}
A_{L,ijk}&=&\frac{e^2}{s_W c_W^2} \sum_{n=1}^4 
m_{\chi_n^0} {\cal C}_0^{a(n,ijk)} Z_N^{1n} 
\left( Z_N^{1n} s_W +  Z_N^{2n} c_W \right) ~, \nonumber \\
C_{L,ijk}&=&-\frac{e^2}{s_W c_W^2} \sum_{n=1}^4 
m_{\chi_n^0} {\cal C}_0^{c(n,ijk)} Z_N^{1n} 
\left( Z_N^{1n} s_W -  Z_N^{2n} c_W \right) ~, \nonumber \\
D_{L,ijk}&=&\frac{e^2}{2 s_W^2 c_W^2} \left\{ \sum_{n=1}^4 
m_{\chi_n^0} {\cal C}_0^{d1(n,ijk)}  
\left( (Z_N^{1n})^2 s_W^2 -  (Z_N^{2n})^2 c_W^2 \right) +
\sum_{m=1}^2 
2 c_W^2 m_{\chi_m} {\cal C}_0^{d2(m,ijk)} Z_{1m}^+ Z_{1m}^- \right\} 
~, \label{FF}
\end{eqnarray}
\end{widetext}

\noindent where $Z_N^{ij}$ and $Z_{ij}^+,~Z_{ij}^-$ 
are the matrices that diagonalize the neutralino and chargino mass
matrices, respectively as defined in \cite{rosiek}, 
$s_W$($c_W$) is the sine(cosine) of the 
weak mixing angle $\theta_W$, and 
${\cal C}_0^{a(n,ijk)},~{\cal C}_0^{c(n,ijk)},~{\cal C}_0^{d1(n,ijk)}$,
${\cal C}_0^{d2(n,ijk)}$ are  
the three-point loop integrals defined via:


\begin{eqnarray}
&&{\cal C}_0 
\left(m_1^2,m_2^2,m_3^2,p_1^2,p_2^2,p_3^2 \right) \equiv \nonumber \\ 
&&\int \frac{d^4 q}{i \pi^2}
\frac{1}{ 
\left[q^2 -m_{1}^2 \right] \left[ (q+p_1)^2 -m_{2}^2 \right]
\left[ (q-p_3)^2 -m_{3}^2 \right] } ~,\\
\end{eqnarray}

\noindent and

\begin{eqnarray}
&&{\cal C}_0^{a(n,ijk)} \equiv {\cal C}_0 
\left(m_{\chi_n^0}^2,m_{{\tilde e}_{kR}}^2,m_{{\tilde e}_{jL}}^2,m_{e_k}^2,
m_{{\tilde \nu}_{iL}}^2,m_{e_j}^2 \right) ~,\nonumber\\
&&{\cal C}_0^{c(n,ijk)} \equiv {\cal C}_0 
\left(m_{\chi_n^0}^2,m_{{\tilde e}_{kR}}^2,m_{{\tilde \nu}_{iL}}^2,m_{e_k}^2,
m_{{\tilde e}_{jL}}^2,m_{\nu_i}^2 \right) ~,\nonumber\\
&&{\cal C}_0^{d1(n,ijk)} \equiv {\cal C}_0 
\left(m_{\chi_n^0}^2,m_{{\tilde \nu}_{iL}}^2,m_{{\tilde e}_{jL}}^2,m_{\nu_i}^2,
m_{{\tilde e}_{kR}}^2,m_{e_j}^2 \right) ~,\nonumber\\
&&{\cal C}_0^{d2(m,ijk)} \equiv {\cal C}_0 
\left(m_{\chi_m}^2,m_{{\tilde e}_{iL}}^2,m_{{\tilde \nu}_{jL}}^2,m_{\nu_i}^2,
m_{{\tilde e}_{kR}}^2,m_{e_j}^2 \right) ~.
\label{integrals}
\end{eqnarray}

\noindent Since the soft RPV trilinear $a$-term in (\ref{slp}) generates 
the soft RPV bilinear term $B_i$ through the RGE [see (\ref{bi})], 
an effective $\tilde l l l$ interaction
($l$ for lepton and $\tilde l$ for slepton) of the type
(\ref{effrpv}) can also arise from slepton--Higgs mixing 
($\propto B_i/B_0$, where $B_0$ is the $\rp$-conserving soft bilinear 
term) followed by 
the Higgs couplings to leptons ($\propto h_l^{jk}$, where $h_l^{jk}$ is the 
Higgs-$l_jl_k$ Yukawa coupling). The contribution of 
this diagram to the $\tilde l l l $ coupling is 
therefore $\propto h_l^{jk} \times B_i/B_0$, and
can be easily estimated when all tree-level low energy SUSY mass 
parameters are assumed to be of the same size, i.e., 
$\sqrt{B_0(M_Z)} \sim \mu(M_Z) 
\sim a_{ijk}(M_Z) \sim M_{SUSY}$. Thus, from (\ref{bi}) 
it is clear that 
$h_l^{jk} \times (B_i/B_0) \sim (h_l^{jk} h_\tau/16 \pi^2) 
\times \ln (M_{Pl}^2/M_Z^2)$. For $j\neq k$ and for 
$j=k=1,2$ this effect is negligible, while for $j=k=3$ (couplings involving 
the $\tau$) we have $\tilde l l_3 l_3 \sim (h_\tau^2/16 \pi^2) 
\times \ln (M_{Pl}^2/M_Z^2)$, which gives $\tilde l l_3 l_3 \sim 10^{-4}$ 
for $\tan^2\beta \sim {\cal O}(10)$ and 
$\tilde l l_3 l_3 \sim 10^{-2}$ if $\tan^2\beta \sim {\cal O}(1000)$.     
This effect will be studied in further detail in \cite{future}.

\noindent In Table \ref{tab1} we give a sample of our numerical results 
for the three effective RPV couplings in (\ref{FF}), corresponding to 
the ``Snowmass 2001'' benchmark points SPS1, SPS2, SPS4 and SPS5 of the 
minimal SUperGRAvity (mSUGRA) 
scenario \cite{snowmass}. The ranges of values in Table \ref{tab1}
for each effective operator are due to slight differences in the slepton 
masses for different generations $i,j,k$.  
We see that the SPS1 and SPS5 scenarios 
give the largest effective couplings, of the order of $10^{-4}-10^{-3}$ 
if $a_{ijk} \sim 16 \pi^2$ GeV $\sim 150$ GeV.   

\begin{table}[htb]
\caption{\label{tab1} Values for the effective  
RPV form factors $A_{L,ijk}$, $C_{L,ijk}$ and $D_{L,ijk}$ 
within the 
Snowmass 2001 benchmark 
points SPS1, SPS2, SPS4 and SPS5 of the mSUGRA parameter space 
\cite{snowmass}.}
\begin{tabular}{c|c|c|c|c}
Effective coupling &SPS1&SPS2&SPS4&SPS5\\
(${\rm GeV}^{-1}$) & & & & \\
\hline
$|A_{L,ijk}| \times 10^4 $ & $3.5-3.6$ & $0.08$  & $0.8-1.3$  & $2.5-2.6$ \\
$|C_{L,ijk}| \times 10^4 $ & $3.4-3.5$ & $0.08$   & $0.7-1.1$ & $2.5-2.6$    \\
$|D_{L,ijk}| \times 10^4 $ & $6.8$    & $0.3$  & $2.0-2.6$ & $5.2-5.3$ \\
\end{tabular}
\end{table}

Let us also evaluate the effective 
RPV form factors $A_{L,ijk}$, $C_{L,ijk}$ and $D_{L,ijk}$ in a low energy 
SUSY framework where the SUSY parameter space is 
defined at the electroweak scale. 
In particular, in the most general case 
$A_{L,ijk}$, $C_{L,ijk}$ and $D_{L,ijk}$ depend on: $\tan\beta$, 
the $\rp$ conserving bilinear Higgs mass term $\mu$, the three 
gaugino mass parameters $M_1,~M_2$ and $M_3=m_{\tilde g}$ 
($m_{\tilde g}$ is the gluino mass) and the slepton masses. 
For simplicity, we assume the GUT relationship between 
the ${\rm SU}(3)$, ${\rm SU}(2)$ and ${\rm U}(1)$ gaugino mass parameters, 
namely, $M_2 = (\alpha_2 / \alpha_3) m_{\tilde g}$ and $M_1 = 
(5 / 3)\tan^2\theta_W M_2$, where $m_{\tilde g}$ is taken to 
be a free parameter. We further assume that all the charged sleptons and 
the sneutrinos get a universal contribution, $m_0$, to their masses at 
the electroweak scale. Thus, the masses of the charged sleptons and the 
sneutrino are given by (including the usual D-term contributions)

\begin{eqnarray}
m^2_{{\tilde e}_R} = m^2_0 - {\sin^{2}{\theta_W}}{m^2_Z}{\cos2\beta} 
\end{eqnarray}
\begin{eqnarray}
m^2_{{\tilde e}_L} = m^2_0 - ({1/2} - {\sin^{2}{\theta_W}})
{m^2_Z}{\cos2\beta}
\end{eqnarray}
\begin{eqnarray}
m^2_{{\tilde \nu}_L} = m^2_0 + \frac{1}{2} m^2_Z \cos2\beta
\end{eqnarray}

\noindent Under these conditions, our relevant low-energy SUSY parameter space 
is reduced to four parameters: $\tan\beta$, $\mu$, $m_{\tilde g}$ and 
$m_0$.  

\begin{figure*}[htb]
\psfull
\begin{center}
\leavevmode
\epsfig{file=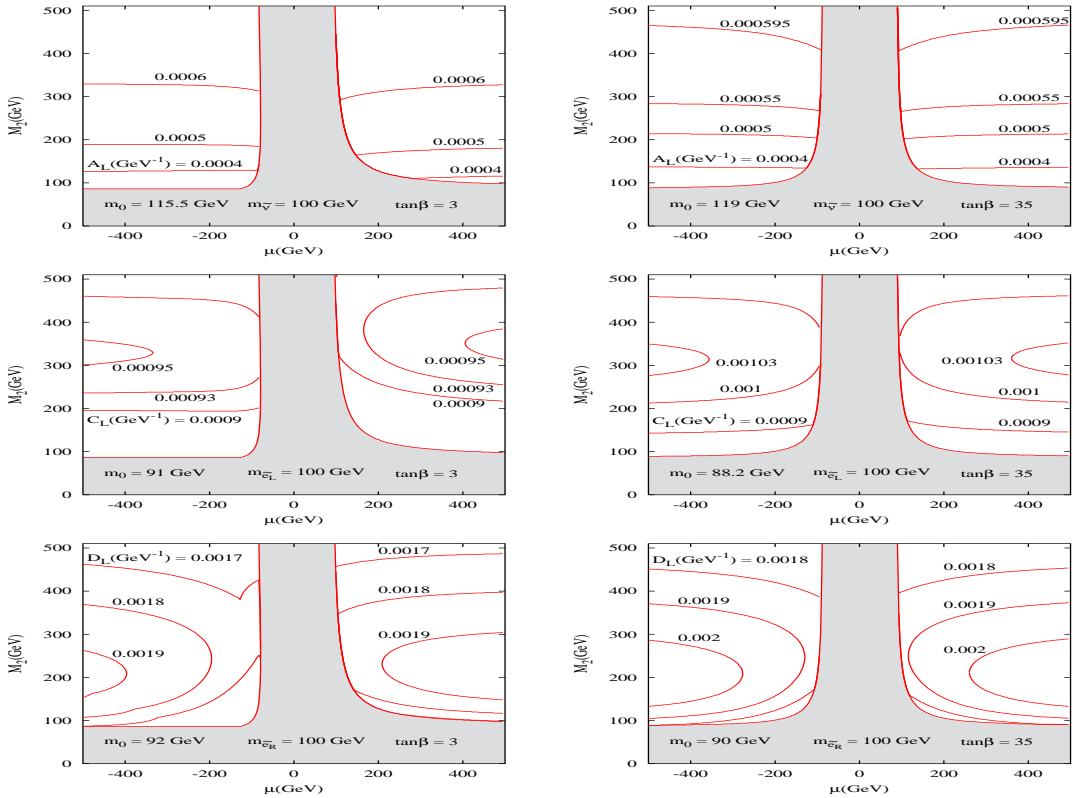,height=12cm,width=20cm,bbllx=0cm,bblly=2cm,
bburx=20cm,bbury=25cm,angle=0}
\end{center}
\caption{\label{fig2} Contours of the RPV effective form factors 
$A_L$, $C_L$ and
$D_L$ (in ${\rm GeV}^{-1}$) in the ($\mu$--$M_2$) plane for $\tan\beta$ = 3 
(left panel) and $\tan\beta$ = 35 (right panel). The shaded region in each 
figure is disallowed by the lighter chargino and the lightest neutralino 
mass bounds from LEP2 $(m_{\chi^\pm_1} > 87$ GeV and $m_{\chi^0_1} > 
23$ GeV) corresponding to the R-parity violating scenario \cite{PDG}. 
See also text.} 
\end{figure*}

Using the above SUSY parameter space, 
in Fig.\ref{fig2} we show contour plots of our 
effective RPV 
couplings $A_L$, $C_L$ and $D_L$ (in ${\rm GeV}^{-1}$) in 
the $\mu-M_2$ plane [$M_2$ is fixed by $m_{\tilde g}$ due to the 
GUT relation mentioned above], 
for two discrete choices of $\tan\beta$, namely, $\tan\beta$ =3 and 35. 
It should be noted that the numbers corresponding to a particular
effective operator are flavor blind since we have assumed a universal 
contribution to the soft scalar masses.
We can see from Fig.\ref{fig2} that for slepton masses of 100 GeV 
(the values of $m_0$ in each of the plots are chosen to give 
slepton masses of 100 GeV) and  
if $a_{ijk} \sim 16 \pi^2$ GeV and $|\mu|$ 
as well as $M_2$ are of the order of several hundreds GeV, then the 
$\tilde\nu ee$, ${\tilde e}_L e \nu$ and ${\tilde e}_R \nu^c e$ effective
couplings in (\ref{effrpv}) range between 
$4 \times 10^{-4} - 2 \times 10^{-3}$, where the ${\tilde e}_R \nu^c e$ 
effective coupling is the largest of the three.

The existing upper
bounds on $\lambda_{ijk}$ (see e.g., Allanach {\it et al.} in \cite{RPVreview}) 
are larger than the expected 
values of our effective couplings
if the trilinear soft breaking RPV coupling, $a_{ijk}$, is of the 
order of the electroweak mass scale.
We note that in the general case of soft lepton number violation, 
the scalar potential may also contain 
the RPV soft operator 
$\epsilon_{ab} a_{ijk}^\prime {\tilde L}^a_i {\tilde Q}^b_j {\tilde D}_k^c$,  
in which case both $\lambda$-like and $\lambda^\prime$-like RPV 
interactions may be effectively loop-generated. Then, the more stringent 
limits on the $\lambda \lambda^\prime$ coupling 
products can be applied to constrain our scenario.
This will be investigated in a future work \cite{future}.  

Finally, in our model the  soft RPV $a$-term in 
(\ref{slp}) gives rise to a neutrino mass only at the two-loop level 
(see \cite{burzumaty}), or equivalently, as an effective ``tree-level" 
neutrino mass which is generated by the sneutrino VEVs ($v_i$) and is 
$\propto v^2_i$. Since $v_i \propto B_i/M_{SUSY}$ \cite{hirsch} and since in 
our model $B_i$ is a one-loop quantity, clearly this also is essentially a 
two-loop effect. Similarly, "one-loop" neutrino masses involving 
two vertices of our effective $\lambda$s are essentially a three-loop
effect.     
  
To summarize, we have postulated that RPV can emerge in the hidden sector 
through the same mechanism that breaks SUSY and then transmitted 
to the SUSY dynamics at some high energy scale, only in   
the form of soft operators. A viable hidden sector model that gives 
rise to softly broken RPV was presented. 

In this scenario Yukawa-like RPV operators are 
generated only through loops involving 
the RPV soft interactions.
Such effective Yukawa-like RPV interactions  
are similar (albeit richer) in structure to the 
conventional $\lambda,~\lambda^\prime$ and $\lambda^{\prime \prime}$ ones 
which, in most RPV scenarios, 
are added ad-hoc in the superpotential.

We have shown that this framework naturally 
explains the smallness and, therefore, the 
present non-observability of RPV interactions in low and high 
energy processes. 
The implications of our effective RPV scenario to the $\lambda^\prime$ and 
$\lambda^{\prime \prime}$-type couplings, as well as to bilinear RPV will be
detailed in \cite{future}.    

We thank J. Wudka, Y. Shadmi, Y. Grossman and S.K. Vempati for very 
helpful discussions. S.R. thanks the Lady Davis Fellowship Trust for 
financial support.

%
%
%


\begin{thebibliography}{99}

\bibitem{RPVreview} See e.g., D.P. Roy, Pramana J. Phys. {\bf 41}, S333 (1993);
G. Bhattacharyya, Nucl. Phys. B (Proc. Suppl.) {\bf 52A}, 83 (1997); P.Roy,
hep-ph/9712520; S. Raychaudhuri, hep-ph/9905576;
B.C. Allanach, A. Dedes and H.K. Dreiner, Phys. Rev. {\bf D60}, 075014 (1999).

\bibitem{burzumaty} F. Borzumati {\it et al.}, Nucl. Phys. 
{\bf B555}, 53 (1999).

\bibitem{related}
See e.g., V. Berezinsky, A.S. Joshipura and J.W.F. Valle, 
Phys. Rev. {\bf D57}, 147 (1998);
J.M. Mira {\it et al.}, Phys. Lett. 
{\bf B492}, 81 (2000); J.L. Chkareuli {\it et al.}, 
Phys. Rev. {\bf D62}, 015014 (2000).

\bibitem{liu} C. Liu and H.S. Song, Nucl. Phys. 
{\bf B545}, 183 (1999).

\bibitem{carlos} See e.g., B. de Carlos and P.L. White, 
Phys. Rev. {\bf D54}, 3427 (1996); A.S. Joshipura, R.D. Vaidya and S.K.
Vempati, Phys. Rev. {\bf D65}, 053018 (2002).

\bibitem{rosiek} J. Rosiek, Phys. Rev. {\bf D41}, 3464 (1990), unpublished
Erratum in hep-ph/9511250.

\bibitem{snowmass} N. Ghodbane and H.-U. Martyn, hep-ph/0201233.

\bibitem{PDG} K. Hagiwara {\it et al.}, Phys. Rev. {\bf D66}, 010001 (2002).

\bibitem{future} S. Bar-Shalom and S. Roy, in preparation.

\bibitem{hirsch} See e.g., M. Hirsch {\it et al.}, Phys. Rev. {\bf D62}, 
113008 (2000); {\it ibid.} {\bf D65}, 119901 (E) (2002); 
A.S. Joshipura and S.K. Vempati, Phys. Rev. {\bf D60}, 111303 (1999);
B. Mukhopadhyaya, S. Roy and F. Vissani, Phys. Lett. {\bf B443}, 191 (1998).


\end{thebibliography}
\end{document}